%% file: main.tex
\documentclass[%
 reprint,
 superscriptaddress,
 amsmath,amssymb,
 aps,
 prc,
]{revtex4-2}

\usepackage{graphicx}
\usepackage{dcolumn}
\usepackage{bm}
\usepackage{hyperref}
\hypersetup{colorlinks,allcolors=black}

\usepackage{threeparttable}

\begin{document}

\title{Lifetime measurement of the muonic atoms of enriched Si isotopes}
\preprint{RIKEN-iTHEMS-Report-25}

\author{R. Mizuno}
\affiliation{Department of Physics, The University of Tokyo, Bunkyo, Tokyo 113-0033, Japan}
\author{M. Niikura}
\email{niikura@ribf.riken.jp}
\affiliation{Nishina Center for Accelerator-Based Science, RIKEN, Wako, Saitama 351-0198, Japan}
\author{S.~Akamatsu}
\affiliation{Department of Physics, Rikkyo University, Toshima, Tokyo 171-8501, Japan}
\author{T.~Fujiie}
\affiliation{Department of Physics, Rikkyo University, Toshima, Tokyo 171-8501, Japan}
\author{K.~Ishida}
\affiliation{High Energy Accelerator Research Organization (KEK), Tsukuba, Ibaraki 305-0801, Japan}
\author{T.~Ito}
\affiliation{Department of Physics, Rikkyo University, Toshima, Tokyo 171-8501, Japan}
\author{T.~Kikuchi}
\affiliation{Department of Physics, Rikkyo University, Toshima, Tokyo 171-8501, Japan}
\author{T.~Matsuzaki}
\affiliation{Nishina Center for Accelerator-Based Science, RIKEN, Wako, Saitama 351-0198, Japan}
\author{F.~Minato}
\affiliation{Department of Physics, Kyushu University, Fukuoka 891-0395, Japan}
\affiliation{Nishina Center for Accelerator-Based Science, RIKEN, Wako, Saitama 351-0198, Japan}
\affiliation{Japan Atomic Energy Agency (JAEA), Tokai, Ibaraki 319-1195, Japan}
\author{J.~Murata}
\affiliation{Department of Physics, Rikkyo University, Toshima, Tokyo 171-8501, Japan}
\author{T.~Naito}
\affiliation{RIKEN Center for Interdisciplinary Theoretical and Mathematical Sciences (iTHEMS), Wako, Saitama 351-0198, Japan}
\affiliation{Department of Physics, The University of Tokyo, Bunkyo, Tokyo 113-0033, Japan}
\author{K.~Shimomura}
\affiliation{High Energy Accelerator Research Organization (KEK), Tsukuba, Ibaraki 305-0801, Japan}
\author{S.~Takeshita}
\affiliation{High Energy Accelerator Research Organization (KEK), Tsukuba, Ibaraki 305-0801, Japan}
\author{I.~Umegaki}
\affiliation{High Energy Accelerator Research Organization (KEK), Tsukuba, Ibaraki 305-0801, Japan}
\author{Y.~Yamaguchi}
\affiliation{Japan Atomic Energy Agency (JAEA), Tokai, Ibaraki 319-1195, Japan}

\date{\today}

\begin{abstract}
\begin{description}
\item[Background] 
A muonic atom, composed of a negative muon and an atomic nucleus, undergoes two primary decay processes: muon decay and muon nuclear capture.
The branching ratio between these two processes can be determined from the measured lifetime of the muonic atom.
While past researches have examined the general trend of muon capture rates across different elements, experimental and theoretical investigations into the isotope dependence of the lifetime remain limited.
\item[Purpose]
The present study aims to measure the lifetimes of the muonic atom of isotopically enriched silicon isotopes.
\item[Methods]
The experiment was conducted at the muon facility in the Material and Life Science Facility (MLF), J-PARC.
A muon beam was stopped in various target materials, including isotopically enriched $^{28,29,30}$Si.
The lifetimes of the muonic atoms were measured by detecting decay electrons using a $\mu$SR spectrometer.
The decay spectra were analyzed by fitting with multi-exponential functions to account for contributions from the target nucleus and surrounding materials.
\item[Results]
For the first time, the lifetimes of the muonic atom of isotopically enriched $^{28,29,30}$Si were measured.
Additionally, seven other targets were studied to validate the experimental method and analysis procedure.
The results were compared with theoretical models such as Primakoff, Goulard-Primakoff, and the recently-developed microscopic and evaporation model (MEM).
\item[Conclusions]
The Primakoff and Goulard-Primakoff formulas, while reproducing certain aspects of the isotope dependence, require further refinement.
The comparison with MEM calculations constrains the axial vector ($g_A$) and induced pseudoscalar ($g_P$) coupling constants.
The present experiment establishes a method for measuring the lifetimes of the muonic atom and will contribute to future systematic investigations.
\end{description}
\end{abstract}

\maketitle

\section{Introduction}
\label{sec:intro}

A muonic atom consists of a negative muon and an atomic nucleus.
The muon at the lowest atomic state ($1s$ state) undergoes either muon decay or muon nuclear capture.
The muon decay, also known as a decay-in-orbit or Michel decay, is a conversion of the muon to an electron, an electron antineutrino, and a muon neutrino, expressed as
\begin{equation}
    \mu^- \rightarrow e^- + \bar{\nu}_e + \nu_\mu.
\end{equation}
Muon nuclear capture~\cite{Measday2001-hi}, also known as ordinary muon capture or muon nuclear absorption, is a conversion of the muon and a proton inside the nucleus to a neutron and a muon neutrino, the elemental process of which is expressed as
\begin{equation}
    \mu^- + p \rightarrow n + \nu_\mu.
\end{equation}

The capture rate of the muon capture reaction varies depending on the nucleus forming the muonic atom.
In general, the capture rate increases for heavier nuclei due to an increased overlap of the muon and nuclear wave functions.
The muon capture rates ($\Lambda_\mathrm{cap}$) are deduced by measurements of a mean lifetime of the muonic atom ($\tau_\mathrm{total}$) as
\begin{equation}
    \frac{1}{\tau_\mathrm{total}} = \Lambda_\mathrm{cap} + \frac{Q}{\tau_{\mu^+}}, \label{eq:lifeconv}
\end{equation}
where $\tau_{\mu^+}$ is the lifetime of the positive muon, viz.~2.1969811(22)~$\mu$s~\cite{pdg2024}, and $Q$ is a Huff factor~\cite{Huff1961-ur}, which takes into account a reduction in the muon decay rate due to the atomic binding energy.
A comprehensive study of the capture rates was conducted, with most results summarized in Ref.~\cite{Suzuki1987-aq}.
However, these measurements focused primarily on natural abundance targets, and data on the capture rates for isotopically enriched targets are still limited.
A significant isotope dependence of the muon nuclear capture rate was pointed out by some measurements in the past~\cite{Bertram1960-fb, Cramer1962-nt, Povel1970-kf, Petitjean1971-xl,  Bobrov1966-xh, Mamedov2000-qz, Fynbo2003-de}.
In general, the capture rate decreases with an increase in the mass number of isotopes, because of the Pauli exclusion principle on the muon capture process~\cite{Primakoff1959-rs}.
As the neutron number increases, more neutron orbitals are filled, making it more difficult for a proton to transform into a neutron at higher levels by muon capture.

Several theoretical calculations have been conducted for the muon nuclear capture rate.
The Primakoff~\cite{Primakoff1959-rs} and Goulard-Primakoff~\cite{Goulard1974-sr} formulas have frequently been used to estimate the capture rate.
Although these formulas provide a good overall approximation, they are less accurate in predicting the isotope dependence of the capture rate, particularly due to the lack of experimental data for isotopically enriched targets~\cite{Iwamoto2025-cf}.
Recently, a microscopic and evaporation model (MEM) was developed to describe the comprehensive muon nuclear capture process, including muon capture, direct and pre-equilibrium processes, and particle evaporations from a compound nucleus~\cite{Minato2023-hr}. 
The calculation reveals a strong dependence of the capture rate on the axial vector and induced pseudoscalar coupling constants ($g_A$ and $g_P$, respectively).
Due to the significant dependence on these constants, the capture rate calculated by MEM also has considerable uncertainty on the capture rate.
Several other theoretical calculations have also been performed~\cite{Chiang1990-vf,Zinner2006-pe,Marketin2009-eu,Giannaka2015-en}.
However, these theories primarily focused on the global systematics of muon capture and did not discuss the detailed isotope dependence.

The present experiment was motivated by our previous measurement of production branching ratios of residual nuclei by muon capture on enriched Si isotopes~\cite{Mizuno2025-ix}.
As the branching ratio of each residual nucleus was given as the number per muon nuclear capture, the capture probability ($P_\mathrm{cap}=\tau_\mathrm{total}\Lambda_\mathrm{cap}$) is necessary to obtain the absolute branching ratio.
The isotope dependence of $P_\mathrm{cap}$ for $^{28,29,30}$Si has not been known to date, and the Primakoff and Goulard-Primakoff formulas do not accurately predict isotope dependence of $P_\mathrm{cap}$~\cite{Bertram1960-fb, Cramer1962-nt, Povel1970-kf, Petitjean1971-xl,  Bobrov1966-xh, Suzuki1987-aq}.
Therefore, in the present experiment, we have measured $P_\mathrm{cap}$, or equivalently, the mean lifetimes for $^{28,29,30}$Si.

\section{Experiment}
The experiment was performed at the D1 area at the muon science facility (MUSE) in the Materials and Life Science Facility (MLF) in J-PARC~\cite{Miyake2012-cr}.
A proton beam from the Rapid Cycling Synchrotron (RCS) with an energy at 3\,GeV irradiated a rotating graphite target to produce pions. 
The primary beam had a double-pulse structure with a 25\,Hz repetition and a power of approximately 900\,kW.
The negative muon beam, a decay product of negative pions, was transported through the D-beamline.
The beam momentum was selected to 32\,$\mathrm{MeV}/c$ to maximize the number of muons stopped at the target and minimize that at the air and target case.
The beam was then delivered to the D1 area, where the experimental apparatus was installed.
The muon beam intensity was approximately $3\times 10^4$~sec$^{-1}$.

\begin{figure}
    \centering
    \includegraphics[width=\linewidth]{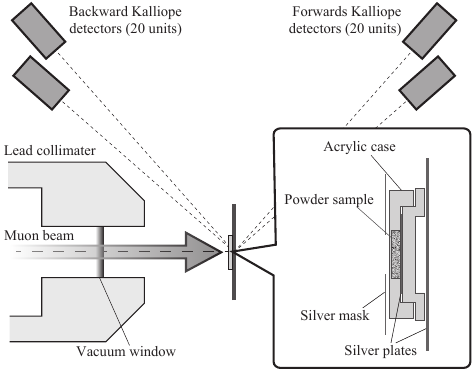}
    \caption{Schematic of the experimental setup at the D1 area in MLF MUSE at J-PARC (not to scale).
    The negative muon beam is delivered from the left side of the figure.
    The beam passes through the lead collimator and stops at the target placed at the center of the $\mu$SR spectrometer.
    A total of 40 units of the Kalliope detectors surrounds the target to detect the decay electrons for the lifetime measurement.
    The detectors are arranged with cylindrical symmetry around the beamline.
    }
    \label{fig:setup}
\end{figure}

Figure~\ref{fig:setup} shows a schematic view of the experimental setup at the D1 area.
The muon beam passed through a lead collimator with a hole size of 40~mm in diameter and irradiated a target.
The targets were located approximately 65 mm downstream of the exit of the lead collimator.
A total of ten targets was measured in the present experiment, as listed in Table~\ref{tab:result}.
The main goal of the present study is to measure the lifetimes of the muonic atom of three isotopically enriched Si targets of $^{28,29,30}$Si. 
The enriched targets had been used in our previous study, and details of the target form, size, weight, chemical purity, and target enrichment are given in Table~\ref{tab:result}.
The targets were in metal chippings form encapsulated in an acrylic case, as schematically illustrated in the inset of Fig.~\ref{fig:setup}.
The effective thickness of the targets was approximately 1.0~g/cm$^2$.
The primary source of uncertainties in the lifetime measurement in the present setup was the muon decay electrons from the light elements of carbon (C), nitrogen (N), and oxygen (O), as discussed in the next section.
The silver plates were used to avoid stopping the muon beam in the acrylic case.
A silver mask with a hole size of 12~mm in diameter and a thickness of 0.5 mm was placed on the upstream side of the case, and a silver backing plate with a thickness of 0.5 mm was installed on the downstream side of the samples.
To evaluate the effect of the chipping form of the target in the acrylic case, natural abundance silicon ($^\mathrm{nat}$Si) samples of a self-supported metal plate and a powder form in the same acrylic case were also measured.
In addition to the Si targets, the lifetimes of the muonic atoms of $^\mathrm{nat}$Mg and $^{27}$Al, which have a similar lifetime to silicon isotopes, were measured to evaluate the reliability of the measurement. 
For $^{27}$Al, both a self-supported plate and a plate in the acrylic case were measured to evaluate the effect of the target case.
In a $^\mathrm{nat}$Ag sample measurement, the lifetimes of background components of the measurements were evaluated.
The $^{55}$Mn target was measured to evaluate the reliability of lifetime measurement in the range between the lifetime of silicon isotopes and silver.
Those targets had an energy loss equivalent thickness of the enriched Si targets and were used to verify the validity of the experimental and analysis procedures.
    
The lifetime of the muonic atom was measured by detecting the decay electrons using a $\mu$SR spectrometer based on Kalliope detectors~\cite{Kojima2014-xa}.
The spectrometer consisted of a total of 40 units of the Kalliope detectors.
Each unit of the Kalliope detectors was comprised of 16 pairs of plastic scintillators.
A coincidence of each pair defined an electron event of the muon decay.
A zero-field correction was applied to avoid the muon spin relaxation and rotation affecting the measured lifetime.
The residual magnetic field at the center of the spectrometer was less than 10\,mG.
The data acquisition system was triggered by the reference timing of the synchrotron.
As the Kalliope detector system can obtain signals until 1 Mcps without saturation~\cite{Wakata2025-ml}, the dead time was negligible in the present experiment with a negative muon beam. 
The timings of the electron event were stored from $-7$ to 57~$\mu$s with respect to the timing of the beam arrival~\cite{Takeshita2024-ck}.
The measurement times for each target are listed in Table~\ref{tab:result}.

\section{Analysis}

\begin{figure}
    \centering
    \includegraphics[width=8.6cm]{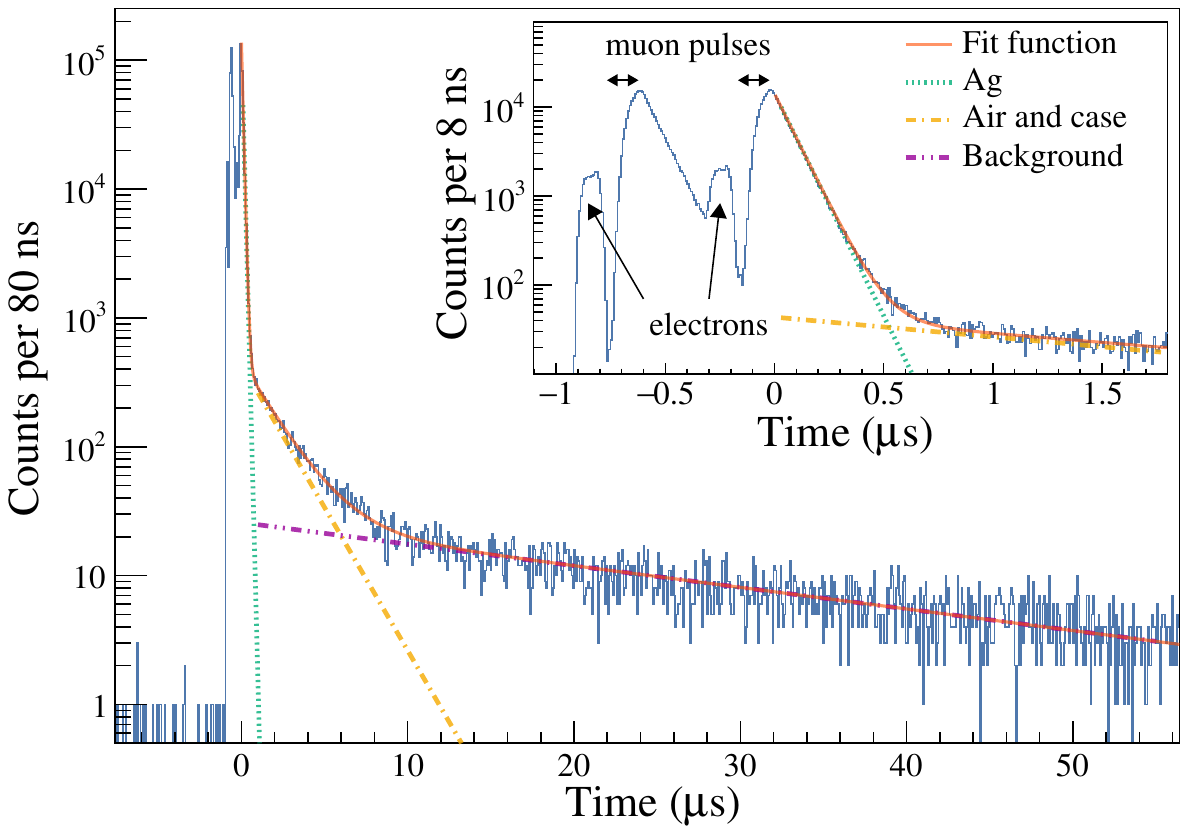}
    \caption{Decay spectrum of a Ag metal plate target in an acrylic case.
    A red solid, a green dotted, a yellow dash-dot, and a purple dash-dot-dot lines represent the fit function, a Ag decay, air and case decay, and a long-lived background, respectively.
    The inset figure is an enlarged view of the same spectrum (with different binning) around the beam irradiation timing.}
    \label{fig:bg}
\end{figure}

\begin{figure*}
    \centering
    \includegraphics[width=18.0cm]{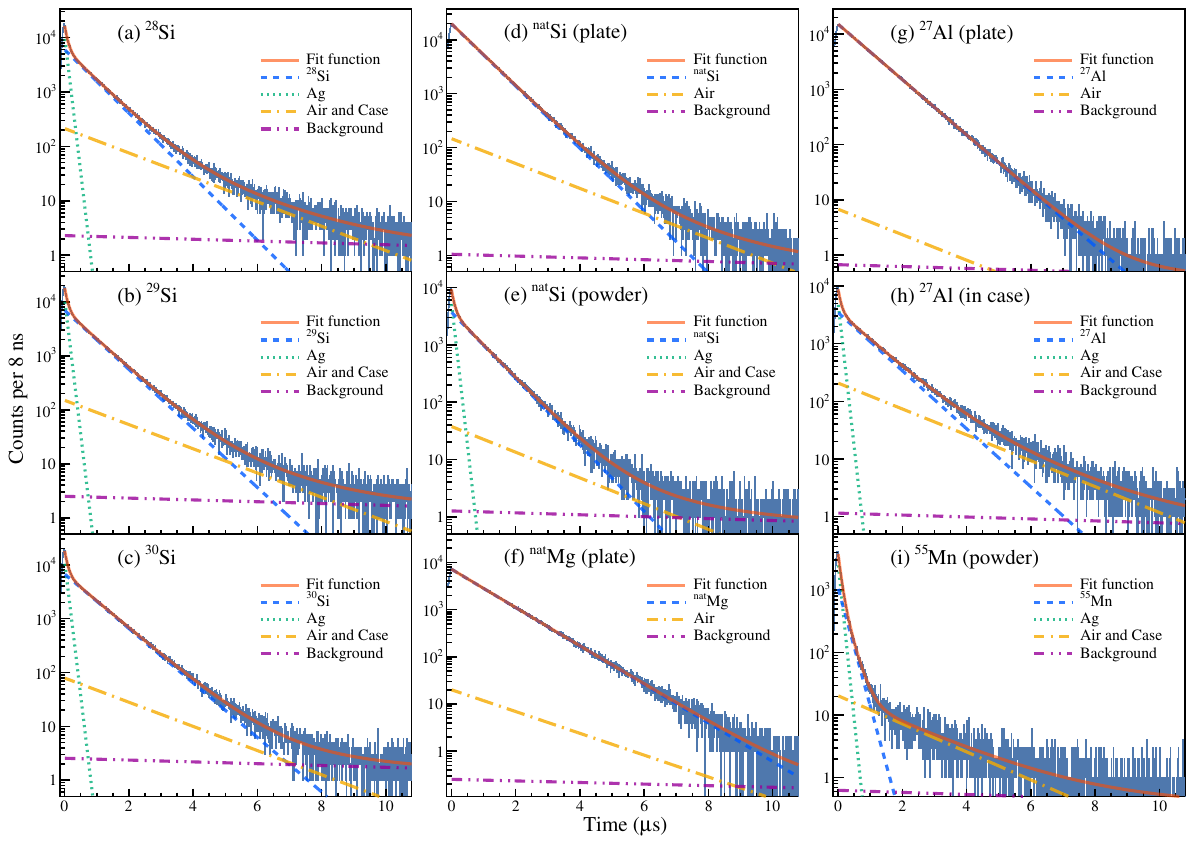}
    \caption{Decay spectra of $^{28,29,30}$Si, $^\mathrm{nat}$Si, $^\mathrm{nat}$Mg, $^{27}$Al, and $^{55}$Mn.
    A red solid, a blue dashed, a green dotted, a yellow dash-dot, and a purple dash-dot-dot lines represent the fit function, a sample decay, a Ag decay, air and case decay, and a long-lived background, respectively.}
    \label{fig:decay}
\end{figure*}

The lifetimes of the muonic atom for each target were deduced by fitting the decay spectra.
These spectra were produced by summing up the timings from all 40 Kalliope detector units.
The timing spectra are shown in Figs.~\ref{fig:bg} and \ref{fig:decay}.

Backgrounds in the spectra were evaluated using an Ag metal plate target in the acrylic case, as shown in Fig.~\ref{fig:bg}.
Due to the double pulse structure of the beam, two peaks were observed in the timing spectra, along with two peaks at an earlier timing than the muon beam, where electrons contaminated the beam scattered at the target and surrounding materials.
The time origin of the spectrum is defined as approximately 40\,ns after the end timing of the second pulse.
The spectrum was fitted by log-likelihood method using the following function consisting of three exponentials:
\begin{equation}
    f_0(t) = a_1 e^{-t/\tau_\mathrm{Ag}}
    + a_2 e^{-t/\tau_\mathrm{CNO}}
    + a_3 e^{-t/\tau_\mathrm{bg}},
    \label{eq:decay0}
\end{equation}
where $\tau_\mathrm{Ag}$ is the lifetime of the $^\mathrm{nat}$Ag muonic atom, $\tau_\mathrm{CNO}$ is the lifetime of the muonic atom of air and the acrylic case, $\tau_\mathrm{bg}$ is a long-lived background, and $a_i$ ($i=1,\,2,\,3$) are coefficients. 
The lifetimes of these three components were well separated and could be deduced by the free fitting.
The second term originated from the events where the muon stopped in the air or the acrylic case, which constitutes approximately 2\% of the total count.
The air and acrylic contain carbon (C), nitrogen (N), and oxygen (O).
Because their lifetimes are similar ($\tau_\mathrm{C}=2026.3(15)$\,ns, $\tau_\mathrm{N}=1906.8(30)$\,ns, and $\tau_\mathrm{O}=1795.4(20)$\,ns~\cite{Suzuki1987-aq}), the decay curve was represented with a single exponential with an \textit{averaged} lifetime ($\tau_\mathrm{CNO}$).
The $\tau_\mathrm{CNO}$ value was determined to be 1.94(8)\,$\mu$s.
The quoted uncertainty contained the statistical uncertainty from the fitting (0.05\,$\mu$s) and the systematic uncertainty. 
The systematic uncertainty originates from variations in the ratio of muons stopped in the air and acrylic case, which depends on the shape of the target.
Hence, an additional systematic uncertainty of 0.03\,$\mu$s was estimated in $\tau_\mathrm{CNO}$, which is used in subsequent analyses.
The third term of Eq.~(\ref{eq:decay0}) represents the long-lived background which was not related to the muon decay.
Although the existence of such backgrounds has been known for a long time, its origin has remained unclear.
It might be attributable to neutrons from the primary beam dump.
The $\tau_\mathrm{bg}$ value was deduced from the fitting to be 25.8(16)\,$\mu$s.
The shape (decay constant) of the long-lived background was independent of the targets, so we used the same value for all targets.
Since the events before the beam arrival were negligible, a constant term was not included in the fitting function.
The lifetime of the muonic Ag was deduced to be 87.0(5)\,ns.

Figure~\ref{fig:decay} shows decay spectra measured with nine targets.
The lifetimes of each target were deduced by fitting the following function:
\begin{equation}
    f_1(t) = a_0 e^{-t/\tau_\mathrm{nucl}} + f_0(t),
    \label{eq:decay1}
\end{equation}
where $\tau_\mathrm{nucl}$ is the lifetime of the muonic atom of the target measured.
Coefficients, $a_i$ ($i=0,\,1,\,2,\,3$), are the parameters to be fitted with $\tau_\mathrm{nucl}$.
If the target was a self-supported metal plate without using the Ag masks and backing plate, the decay component of Ag was omitted in the fitting, i.e., $a_1=0$.
For the self-supported metal targets, $\tau_\mathrm{CNO}$ was deduced independently from the Ag sample measurement as discussed above. 
The primary background contribution, the second term of Eq.~(\ref{eq:decay0}), was attributed to air, which contains less carbon. 
The lifetime of 1.88(5)\,$\mu$s was used based on the composition of air and the muonic lifetime of nitrogen and oxygen. 
The muon stopping probability in the air was estimated to be approximately 1\% using the GEANT4 toolkit~\cite{geant4-1,geant4-2,geant4-3}.
This value was consistent with the stopping rate deduced from fits to data obtained with the self-supported metal targets.
Note that the hyperfine splitting of muon $1s$ state of $^{27}$Al can affect the lifetime of the muonic $^{27}$Al.
Nevertheless, this effect was not considered; the decay curve was fitted with a single exponential function following the same analysis procedure as in Ref.~\cite{Suzuki1987-aq}. 

The uncertainty in the lifetimes was estimated using the Monte Carlo method.
In this method, the fixed parameters ($\tau_\mathrm{Ag}$, $\tau_\mathrm{CNO}$, and $\tau_\mathrm{bg}$) and fitting range were randomly varied within their respective uncertainties.
The timing spectrum was then fit 500 times, resulting in a distribution of $\tau_\mathrm{nucl}$ values.
The standard deviation of this distribution was taken as the final uncertainty in $\tau_\mathrm{nucl}$.
Table~\ref{tab:result} summarizes the obtained lifetimes for each target.

\section{Results}

\input{table_result}

\begin{figure}
    \centering
    \includegraphics[width=8.6cm]{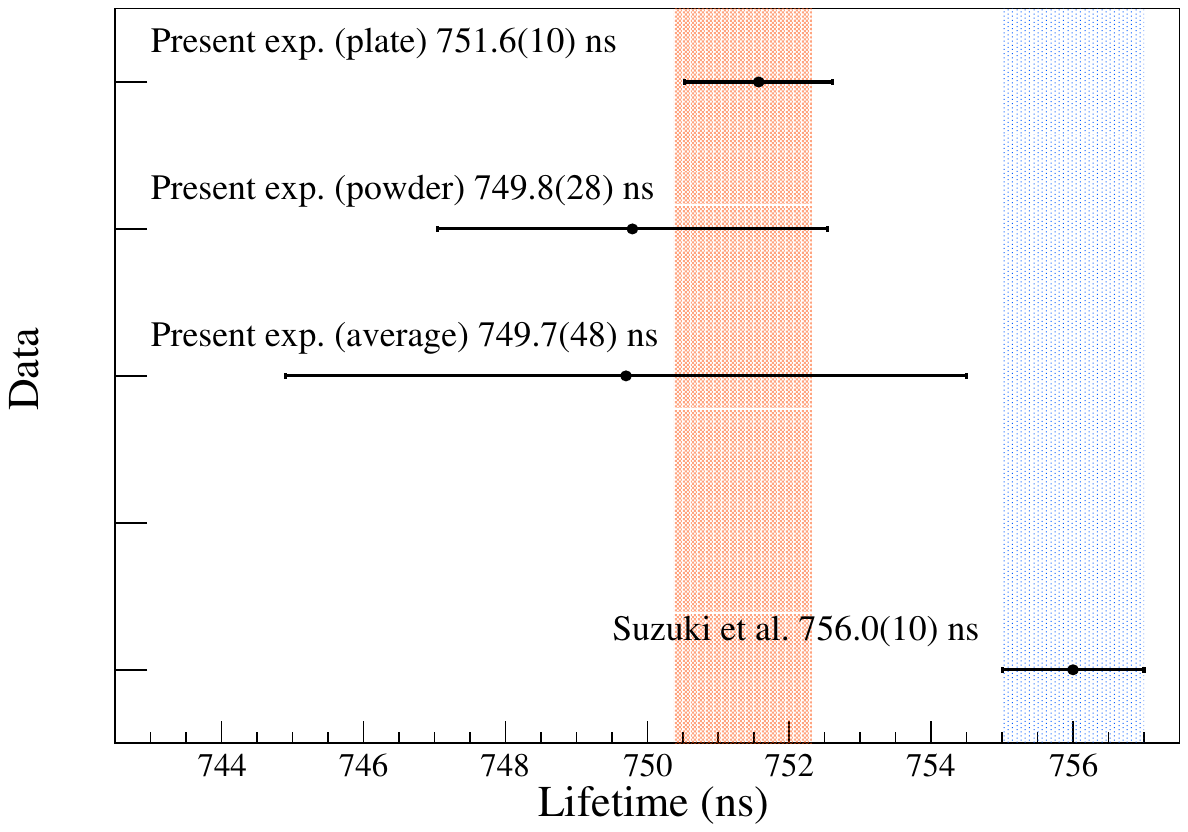}
    \caption{Comparisons of the lifetime of $^\mathrm{nat}$Si measured in the present study and previous one~\cite{Suzuki1987-aq}. The red shaded area indicates the adopted value of 751.4(10)\,ns.}
    \label{fig:comp}
\end{figure}

Table~\ref{tab:result} summarized the measured lifetime and corresponding capture rates, calculated using Eq.~(\ref{eq:lifeconv}).
The Huff factors used in the conversion were taken from Ref.~\cite{Suzuki1987-aq} and are included in the table.
In this study, the lifetimes of isotopically enriched $^{28,29,30}$Si were obtained for the first time.
The lifetimes of $^\mathrm{nat}$Mg, $^\mathrm{27}$Al, $^\mathrm{55}$Mn, and $^\mathrm{nat}$Ag were found to be consistent with previous measurement~\cite{Suzuki1987-aq}, confirming the reliability of our experimental setup and data analysis.
However, the lifetime of $^\mathrm{nat}$Si deviated from the previously reported value.
To investigate the effect of the different forms of the targets and the acrylic case, we measured two $^\mathrm{nat}$Si targets: a self-supported metal plate and a powder form encapsulated in the acrylic case. 
Both targets yielded consistent results, confirming that the shapes of the targets did not influence the measured lifetime.
Additionally, the lifetime of $^\mathrm{nat}$Si was also deduced from that of the enriched targets obtained in the present study.
A Monte-Carlo calculation was used to create the decay curve of $^\mathrm{nat}$Si by summing the timing spectrum of enriched Si, weighted by the natural abundance~\cite{natural_abundance}.
By fitting this simulated decay curve, the average lifetime of $^\mathrm{nat}$Si was deduced to be 749.7(48)\, ns, consistent with our measurements as illustrated in Fig.~\ref{fig:comp}.
While our adopted value of 751.4(10)\,ns, an uncertainty-weighted average of the present results, deviates from the previous value of 756.0(10)\,ns, the reason for this discrepancy remains unclear.

\section{Discussions}

\input{table_err}

In this study, we have obtained the muon nuclear capture rates of ten targets.
As the present study was the first investigation of muonic lifetime using the $\mu$SR spectrometer at MLF, we first discuss the measurement precision.
Table~\ref{tab:err} details the systematic breakdown of the uncertainty.
Propagated uncertainties from each parameter were estimated using the Monte-Carlo method by keeping other parameters fixed.
The statistical uncertainty on $\tau_\mathrm{nucl}$ for most targets was approximately 1--2\,ns.
For targets in acrylic cases, the propagated uncertainty from $\tau_\mathrm{CNO}$ dominated the total uncertainty.
For the self-supported metal plate targets, namely $^\mathrm{nat}$Si, $^\mathrm{nat}$Mg, and $^{27}$Al, the statistical uncertainty dominated the total uncertainty, with uncertainties propagated from $\tau_\mathrm{CNO}$ and $\tau_\mathrm{bg}$ having a minor impact on the overall precision.
The present results demonstrate the potential of our method to measure the lifetimes of various nuclei with a precision of 1--2\,ns within a few hours of measurement.

\begin{figure}
    \centering
    \includegraphics[width=8.6cm]{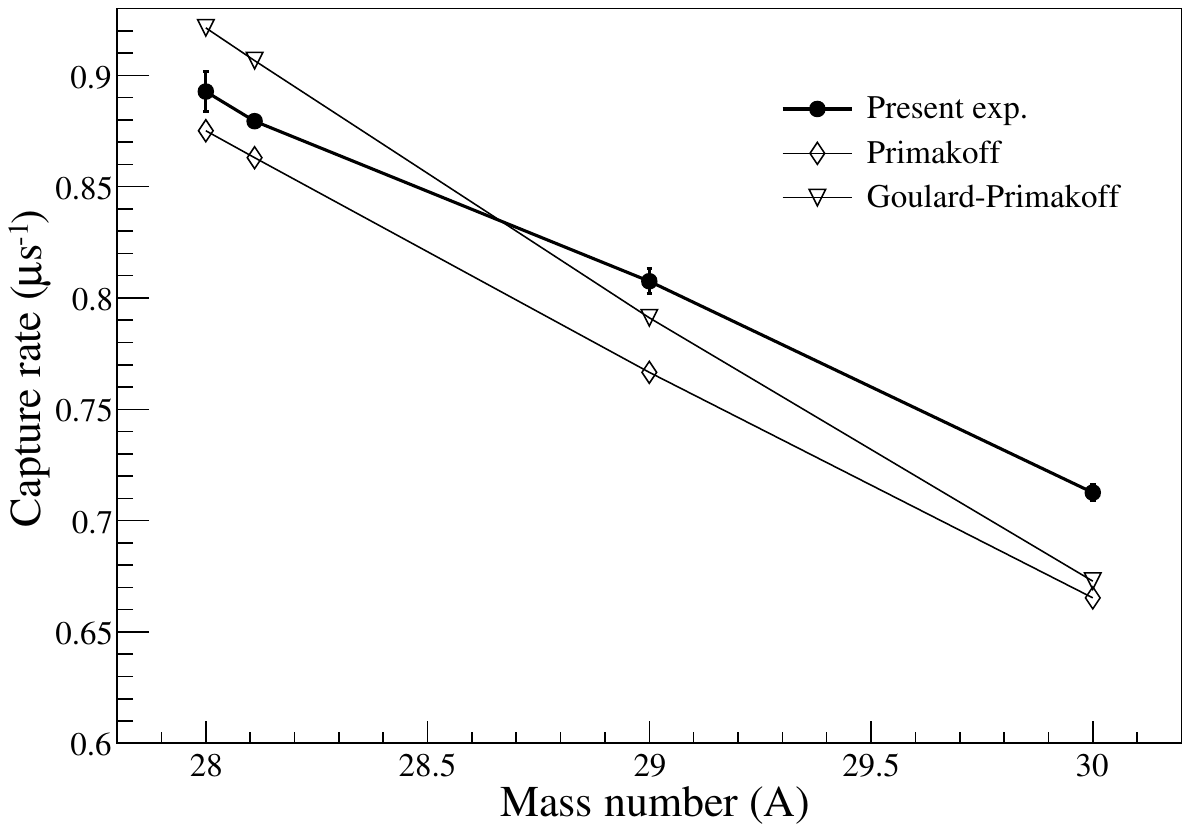}
    \caption{Isotope dependence of the capture rates for muonic Si comparing with theoretical calculations. 
    Closed circles represent the experimental values from the present study, and open symbols are theoretical calculations. The data for $^\mathrm{nat}$Si is also plotted at an average mass number of 28.11.}
    \label{fig:si}
\end{figure}

Table~\ref{tab:comp} compares the capture rate obtained in the present study with theoretical calculations.
The capture rates for $^\mathrm{nat}$Si and $^{27}$Al in the table were calculated by taking an uncertainty-weighted average of two experimental data.
The isotope dependence of the capture rate for the silicon isotopes is illustrated in Fig.~\ref{fig:si}.
The data for $^\mathrm{nat}$Si is also plotted at an average mass number of 28.11.
The experimental capture rates decrease linearly with increasing mass numbers.
This constant rate of decrease indicates that the capture rate is not influenced by the odd-even effect of the neutron number.

The isotope dependence of the capture rate was compared with the Primakoff~\cite{Primakoff1959-rs} and Goulard-Primakoff~\cite{Goulard1974-sr} formulas, which are currently the only theoretical expressions available for comparison with experimental data in isotope dependence.
The Primakoff formula is given by
\begin{equation}
    \Lambda_\mathrm{cap}(A,Z)
    = Z_\mathrm{eff}^4X_1\left[1-X_2 \left( \frac{A-Z}{2A} \right)\right],
    \label{eq:primakoff}
\end{equation}
where $Z_\mathrm{eff}$ represents an overlap of the wave functions between the muon and nucleus, and $A$ and $Z$ are the mass number and the atomic number of the target nucleus, respectively.
While most previous studies and applications~\cite{Suzuki1987-aq,Abe2017-xw,Battistoni2015-ht,geant4-3,Iwamoto2025-cf} have taken the $Z_\mathrm{eff}$ values from original work by Ford and Wills~\cite{Ford1962-qu}, we use $Z_\mathrm{eff}$ obtained by a theoretical calculation explained below.
The parameters $X_1$ and $X_2$ are coefficients obtained from a global fit of the experimental data, where $X_1=170\,s^{-1}$ and $X_2=3.125$ taken from Ref.~\cite{Suzuki1987-aq} were used in the following calculation.
Equation~(\ref{eq:primakoff}) was later modified to introduce higher order corrections by Goulard and Primakoff~\cite{Goulard1974-sr} as 
\begin{align}
    \Lambda_\mathrm{cap}(A,Z) 
    = Z_\mathrm{eff}^4G_1 \left[ 1+G_2\frac{A}{2Z}
      -G_3\frac{A-2Z}{2Z} \right. \notag \\
      \left.
      -G_4\left(
      \frac{A-Z}{2A}+\frac{A-2Z}{8AZ}
      \right)
      \right],
      \label{eq:goulard-primakoff}
\end{align}
where the parameters $G_1=261\,s^{-1}$, $G_2=-0.040$, $G_3=-0.26$, and $G_4=3.24$ were taken from TRIUMF data~\cite{Suzuki1987-aq}.

The capture rate is predominantly influenced by two effects: an overlap of the wave functions between the muon and the nucleus, and the Pauli exclusion effect, which is expressed as $Z_\mathrm{eff}$, and parameters of $X_2$ and $G_{2,3,4}$ in the Primakoff and Goulard-Primakoff formulas, respectively.

To investigate the isotope effect on $Z_\mathrm{eff}$, we performed a microscopic calculation using the definition proposed by Ford and Wills~\cite{Ford1962-qu} as
\begin{equation}
  Z_{\mathrm{eff}}^4 = \pi a_{\mathrm{B}}^3
  \int_0^{\infty}
  \left\{
    \left[ F \left( r \right) \right]^2
    +
    \left[ G \left( r \right) \right]^2
  \right\}
  \rho_{\mathrm{ch}} \left( r \right) dr,
  \label{eq:zeff}
\end{equation}
where $a_{\mathrm{B}} = 2.559 \times 10^2$\,fm is the Bohr radius for muons~\cite{Tiesinga2021Rev.Mod.Phys.93_025010}, $\rho_{\mathrm{ch}} \left( r \right) $ is the charge density distribution of the nucleus, and $F \left( r \right) $ and $ G \left( r \right)$ are the radial component of the upper and lower components of the Dirac spinor of the muon in the $1s$ orbital, respectively.
Note that the normalization of the Dirac spinor is
\begin{equation}
  \label{eq:normalization}
  \int_0^{\infty}
  \left\{
    \left[ F \left( r \right) \right]^2
    +
    \left[ G \left( r \right) \right]^2
  \right\}
  dr
  =
  1.
\end{equation}
The charge density distribution $\rho_{\mathrm{ch}} \left( r \right)$ was calculated by using the Skyrme Hartree-Fock-Bogoliubov calculation~\cite{Vautherin1972Phys.Rev.C5_626,Dobaczewski1984Nucl.Phys.A422_103} with assuming the spherical symmetry, where $\rho_{\mathrm{ch}} \left( r \right)$ of odd-mass nuclei was calculated using the equal filling approximation without considering the blocking or the time-odd component.
The SGII effective interaction~\cite{sg2} and the volume-type pairing with the 60\,MeV cutoff in the Hartree-Fock equivalent energy~\cite{Stoitsov2013Comput.Phys.Commun.184_1592} were used for the particle-hole and particle-particle channel, respectively~\cite{Naito2023Phys.Rev.C107_054307}.
The Coulomb potential formed by $ \rho_{\mathrm{ch}} \left( r \right)$, the vacuum polarization formed by the point charge~\cite{Uehling1935Phys.Rev.48_55, WayneFullerton1976Phys.Rev.A13_1283}, and the reduced mass calculated based on the AME2020~\cite{Huang2021Chin.Phys.C45_030002,Wang2021Chin.Phys.C45_030003} were considered for $ F \left( r \right) $ and $ G \left( r \right) $~\cite{Minato2023-hr}.

\input{table_comp}

Table~\ref{tab:comp} lists the calculated $Z_\mathrm{eff}$ values.
For the silicon isotopes, the calculated $Z_\mathrm{eff}$ is slightly larger than the original value of $Z_\mathrm{eff}=12.22$~\cite{Ford1962-qu}.
Because this constant value was used to deduce parameters of $X_{1,2}$ and $G_{1,2,3,4}$ in the Primakoff and Goulard-Primakoff formulas~\cite{Suzuki1987-aq}, respectively, absolute values of the calculated capture rates by these formulas may be changed by reevaluating the parameters using our $Z_\mathrm{eff}$ values.
Thus, the isotope dependence of the capture rate is discussed with relative values normalized to that of $^{28}$Si.
While the reductions of $\Lambda_\mathrm{cap}$ are 9.54(12)\% and 20.18(23)\% for $^{29}$Si and $^{30}$Si, respectively, the corresponding reductions in $Z_\mathrm{eff}^4$ are two order of magnitude smaller of 0.10\% and 0.22\%.
Thus, the effect of $Z_\mathrm{eff}$ on the isotope dependence of $\Lambda_\mathrm{cap}$ is negligible for the Si isotopes, in contrast to an Eu isotope case~\cite{Petitjean1971-xl}.
The significant isotope effect from $Z_\mathrm{eff}$ between $^{151}$Eu$_{88}$ and $^{153}$Eu$_{90}$ might be due to the large change in nuclear deformation in the region of the neutron number from 82 to 92.
In the case of the Si isotopes, the quadrupole deformation parameters ($\beta_2$) were experimentally deduced from the $E2$ transition probability between $2^+_1$ and $0^+_\mathrm{gs}$ to be 0.407(7) and 0.315(7) for $^{28}$Si and $^{30}$Si, respectively~\cite{Raman2001-zx}.
The nuclear deformation affects $Z_\mathrm{eff}$ through a root-mean-squere charge radius ($r_\mathrm{rms}$) of $\rho_\mathrm{ch}$ through Eq.~(\ref{eq:zeff}).
The experimenatal $r_\mathrm{rms}$ were 3.1224(24) and 3.1336(40)~fm in Ref.~\cite{Angeli2013-nx} and our calculated $r_\mathrm{rms}$ of $\rho_\mathrm{ch}$ were 3.15 and 3.16\,fm for $^{28}$Si and $^{30}$Si, respectively.
The calculated $r_\mathrm{rms}$ is consistent with the experimental values, although the calculation assumed spherical symmetry.
Therefore, we conclude that the isotope dependence of $\Lambda_\mathrm{cap}$ for the Si isotopes is not explained by that of $Z_\mathrm{eff}$.

The Pauli exclusion effect was expressed in the term with the parameter $X_2$ and $G_{2,3,4}$ in the Primakoff and Goulard-Primakoff formulas, respectively.
While the absolute capture rates calculated by the Primakoff formula underpredict experimental data, the isotope dependence is reasonably well reproduced, as shown in Fig.~\ref{fig:si}.
The Goulard-Primakoff formula overestimates the capture rates of $^{28}$Si and $^\mathrm{nat}$Si, while it underestimates those of $^{29}$Si and $^{30}$Si.
The inadequacy of both Primakoff and Goulard-Primakoff formulas in accounting for the isotope effect has also been previously pointed out in Refs.~\cite{Suzuki1987-aq,Iwamoto2025-cf}.
A re-evaluation of these Primakoff's formulas is a worthwhile direction for future research, and the experimental method established in this study offers an opportunity for systematic measurements.

Capture rates obtained by other theories~\cite{Chiang1990-vf,Zinner2006-pe,Marketin2009-eu,Giannaka2015-en} are also listed in Table~\ref{tab:comp}.
The global systematics of the experimental capture rates and comparisons with those theories are out of the scope of the present study and are discussed in a separated publication~\cite{Iwamoto2025-cf}.

The measured isotope dependence of the muon capture rate was also compared with a microscopic and evaporation model (MEM) calculation~\cite{Minato2023-hr}.
In this model, the wave function of the muonic atom was calculated using the same method as that used to calculate $Z_\mathrm{eff}$.
The muon capture rates were calculated using the second Tamm-Dancoff approximation~\cite{Minato2016-qm} with effective interactions of SkO'~\cite{sko2} and SGII~\cite{sg2}.
Calculated capture rates strongly depend on the axial vector and induced pseudo-scalar coupling constants ($g_A$ and $g_P$, respectively), as well as the choice of effective interaction.
As the capture rate calculated by MEM has large ambiguity due to the uncertainties in $g_A$ and $g_P$ in medium, possible combinations of the parameter set of $g_A$ and $g_P$ were deduced to reproduce the measured capture rates.
Note that the values of $g_A$ and $g_P$ presented in this study are at zero momentum transfer, namely $g_A(q^2=0)$ and $g_P(q^2=0)$, respectively.
Due to the limitation of the MEM calculation to even-even nuclei, we discuss $^{28}$Si and $^{30}$Si hereinafter.

\begin{figure}
    \centering
    \includegraphics[width=8.6cm]{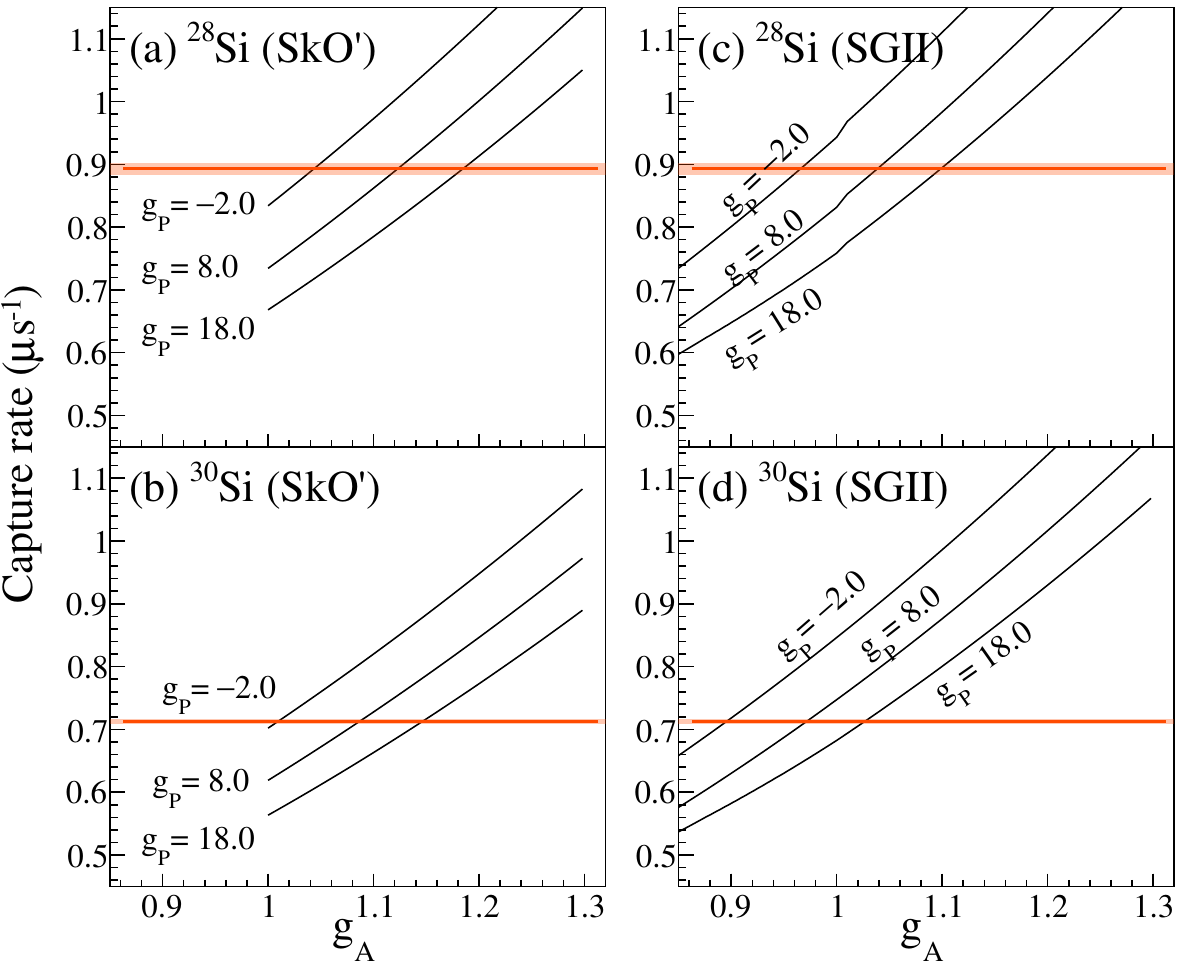}
    \caption{The variation of the muon capture rate with changing $g_A$ and $g_P$ for $^{28}$Si and $^{30}$Si using SkO' and SGII effective interactions.
    The black-solid lines represent the calculated capture rates for a given $g_P$ value as functions of $g_A$.
    The red-solid lines and red-shaded areas represent the experimental capture rates and their 1$\sigma$ uncertainties, respectively.}
    \label{fig:gA_gP_dependence}
\end{figure}

\begin{figure}
    \centering
    \includegraphics[width=8.6cm]{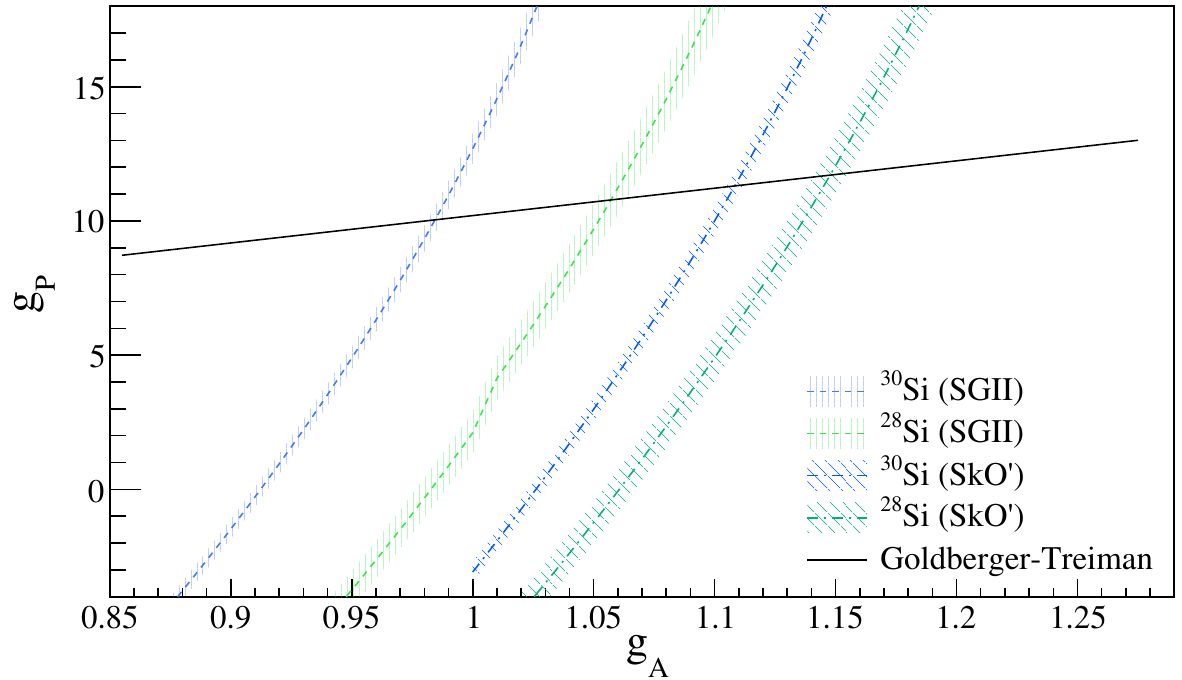}    
    \caption{The parameter sets of $g_A$ and $g_P$, which reproduce the experimental muon capture rate for $^{28}$Si and $^{30}$Si using SkO' and SGII effective interactions.
    The shaded areas represent the 1$\sigma$ uncertainties of the experimental values.
    The black-solid line shows the Goldberger-Treiman relation of $g_P/g_A=10.2$.}
    \label{fig:gAgP_reproduce}
\end{figure}

Figure~\ref{fig:gA_gP_dependence} shows a variation of the muon capture rates by changing $g_A$ and $g_P$ using the SKO' and SGII effective interactions.
The solid lines represent the calculated capture rates for a given $g_P$ value as functions of $g_A$.
The range of $g_A$ was set to 1.0--1.3 and 0.7--1.3 for SkO' and SGII, respectively, which include the value of free nucleon ($g_A=1.27$) and typical quenched value ($g_A=1.0$)~\cite{suhonen_review}.
The range of $g_P$ was set to cover the range to reproduce the experimental result with the given $g_A$ range, including the free nucleon value with the muon capture on the hydrogen of $g_P(q^2=0)=12.1(8)$~\cite{Andreev2013-xo} and other previous investigations~\cite{Jokiniemi2019143, Measday2001-hi, Gorringe2003-gc}.
The muon capture rates are raised with increasing $g_A$ and decrease with increasing $g_P$ in all cases.
The experimental results of capture rates are also shown in Fig.~\ref{fig:gA_gP_dependence} with the red lines.
By taking an overlap between the experimental values and calculated ones, parameter sets of $g_A$ and $g_P$, which reproduce the experimental muon capture rates, were obtained, as illustrated in Fig.~\ref{fig:gAgP_reproduce}.
In the figure, the lines represent the ($g_A$, $g_P$) set constrained by experimental results, and the 1$\sigma$ uncertainties of the experimental values are shown with colored shade.
The Goldberger-Treiman relation, which expresses the relation between $g_A$ and $g_P$ based on the assumption of Partially Conserved Axial Current (PCAC) and the pion-pole dominance of the induced pseudoscalar form factor, yields $g_P(0)/g_A(0)=10.2$~\cite{Goldberger1958-xd}, as shown by the black solid line in the figure.

The $g_A$ values smaller than that of free nucleon of 1.27 are required to reproduce the experimental results with the given range of $g_P$, suggesting the quenching of $g_A$.
When comparing the parameter set between $^{28}$Si and $^{30}$Si with the same effective interactions, there were no overlapping of parameter sets, indicating the isotope dependence of quenching on both or either of $g_A$ and $g_P$.
The larger quenching of $g_A$ was required for $^{30}$Si than $^{28}$Si when $g_P$ was fixed, consistent with the trend discussed in Refs.~\cite{Fearing1992-sj, Gorringe1998-zc, Barea2013-nf, suhonen_review}.
A comparison of constrained ($g_A$, $g_P$) regions for two effective interactions, SkO' and SGII, revealed that the calculation using SkO' reproduced the capture rate with smaller quenching than that using SGII.
This discrepancy can be understood by different energy distributions populated by muon capture.
The excited states calculated with SkO' distribute at higher energies compared to SGII, and phase spaces available for muon neutrinos in the final state are limited.
This results in the smaller muon capture rate of SkO' than SGII~\cite{Minato2016-qm}, leading to the smaller quenching.

\section{Conclusion}
We have measured the lifetimes of the muonic atom using a $\mu$SR spectrometer at the muon facility in MLF, J-PARC.
In this study, the lifetimes of isotopically enriched $^{28,29,30}$Si were obtained for the first time.
The lifetimes of $^\mathrm{nat}$Mg, $^\mathrm{27}$Al, $^\mathrm{55}$Mn, and $^\mathrm{nat}$Ag were also measured and found to be consistent with previous measurements, confirming the reliability of the experimental method and analysis procedure.

We compared the isotope dependence of the lifetimes for $^{28,29,30}$Si with the Primakoff and Goulard-Primakoff formulas, utilizing newly calculated $Z_\mathrm{eff}$ values.
The isotope dependence of $Z_\mathrm{eff}$ is negligibly small and does not explain that of the capture rate.
While the absolute lifetimes calculated by the Primakoff formula exceed our experimental data, the isotope dependence is reasonably well-reproduced.
The Goulard-Primakoff formula accurately predicts the lifetimes of $^{28}$Si and $^\mathrm{nat}$Si but fails to reproduce those of $^{29}$Si and $^{30}$Si.

Through a comparison of our experimental results on $^{28,30}$Si with the recent microscopic and evaporation model (MEM) using two effective interactions (SkO' and SGII), we constrained the axial vector ($g_A$) and induced pseudoscalar ($g_P$) coupling constants of the weak interaction.
The constrained values of $g_A$ are found to be smaller than the free nucleon value of 1.27, suggesting the quenching of $g_A$ in the MEM calculation.
The amount of quenching exhibits isotope dependence between $^{28}$Si and $^{30}$Si.
Furthermore, the choice of effective interaction also influences the required quenching of $g_A$ and $g_P$.
  
The present experiment establishes a method for measuring the lifetimes of the muonic atom at MLF in J-PARC, opening up a new possibility for systematic investigations of capture rates.
The re-evaluation of the muon capture rates with the Primakoff and Goulard-Primakoff formulas will provide better estimation in general-purpose simulation codes such as GEANT4~\cite{geant4-3}, PHITS~\cite{Abe2017-xw}, and FLUKA~\cite{Battistoni2015-ht}.
The systematic study of the capture rate will contribute to a further understanding of the quenching of $g_A$ and $g_P$ of the nuclear medium.

\begin{acknowledgments}
We are grateful for the valuable discussions with Dr.~T.~Y.~Saito (RIKEN), Dr.~H.~Iwamoto (JAEA), and Dr.~T.~Miyagi (Univ.~Tsukuba).
The muon experiment at the Materials and Life Science Experimental Facility of the J-PARC was performed under a user program (Proposal No.~2024MI01).
The present work is partially supported by the JSPS Grant-in-Aid under Grant Nos.~JP19H05664, JP22K20372, JP23H04526, JP23H01845, JP23K01845, JP23K03426, JP24H00073, JP24K00647, and JP24K17057.
R.M. is supported by the Force-front Physics and Mathematics Program in Drive Transformation (FoPM), a World-leading Innovative Graduate Study (WINGS) program, and the JSR Fellowship from the University of Tokyo.
T.N. acknowledges
the RIKEN Special Postdoctoral Researcher Program.
The numerical calculations were partially performed on cluster computers at the RIKEN iTHEMS program.

\end{acknowledgments}

\bibliography{paperpile,naito,gAgP}

\end{document}

%% file: table_result.tex
\begin{table*}
    \centering
    \caption{List of the target nucleus, target form, size and weight, measurement time, Huff factor ($Q$) taken from Ref~\cite{Suzuki1987-aq}, measured lifetime of the muonic atom ($\tau_\mathrm{nucl}$), and capture rate ($\Lambda_\mathrm{cap}$).}
    \label{tab:result}
    \begin{ruledtabular}
    \begin{tabular}{ccccD{.}{.}{5} D{.}{.}{5} D{.}{.}{5} D{.}{.}{6} D{.}{.}{12}}
    \multicolumn{4}{c}{Target}&
    \multicolumn{1}{c}{Time (h)}&
    \multicolumn{1}{c}{Huff factor} &
    \multicolumn{2}{c}{Lifetime (ns)} &
    \multicolumn{1}{c}{Capture rate ($\mu$s$^{-1}$)}\\

    Nucleus & Form & Size (mm) & Weight (g) &
    \multicolumn{1}{c}{}&
    \multicolumn{1}{c}{}&
    \multicolumn{1}{c}{Present}&
    \multicolumn{1}{c}{Suzuki \textit{et al.}\cite{Suzuki1987-aq}}&
    \multicolumn{1}{c}{Present}
    \\\hline

    $^{28}$Si         & Powder in case       & $\phi$15.0$\times$2.8  & 0.500 & 2.02 & 0.992 & 743.9(50)  &            & 0.893(9)   \\ 
    $^{29}$Si         & Powder in case       & $\phi$15.0$\times$2.8  & 0.500 & 2.05 & 0.992 & 794.2(35)  &            & 0.808(6)   \\ 
    $^{30}$Si         & Small pieces in case & $\phi$15.0$\times$2.8  & 0.500 & 2.10 & 0.992 & 859.0(27)  &            & 0.713(4)   \\ 
    $^\mathrm{nat}$Si & Self-supported plate & $\phi$50$\times$1.96   & 9.00  & 1.45 & 0.992 & 751.6(10)  & 756.0(10)  & 0.8790(18) \\ 
    $^\mathrm{nat}$Si & Powder in case       & $\phi$15.0$\times$2.8  & 0.550 & 1.06 & 0.992 & 749.8(28)  & 756.0(10)  & 0.882(5)   \\ 
    $^\mathrm{nat}$Mg & Self-supported plate & 50$\times$50$\times$2.0& 8.54  & 0.43 & 0.995 & 1065.5(20) & 1067.2(20) & 0.4856(18) \\
    $^{27}$Al         & Self-supported plate & 50$\times$50$\times$2.0& 13.6  & 0.98 & 0.993 & 863.7(10)  & 864.0(10)  & 0.7059(13) \\ 
    $^{27}$Al         & Plate in case        & $\phi$14.8$\times$1.0  & 0.451 & 0.94 & 0.993 & 862.5(81)  & 864.0(10)  & 0.707(11)  \\ 
    $^{55}$Mn         & Chippings in case    & $\phi$15.0$\times$2.0  & 2.127 & 0.49 & 0.976 & 230.3(45)  & 232.5(20)  & 3.90(8)   \\ 
    $^\mathrm{nat}$Ag & Plate in case        & $\phi$14.8$\times$0.3  & 0.544 & 2.87 & 0.925 &  87.0(5)   & 87.0(15)   & 11.07(7)   \\ 
    \end{tabular}
    \footnotetext[0]{Chemical purity and isotope enrichment of the $^{28}$Si target are 99.3218\% and 99.93\%, respectively.}
    \footnotetext[0]{Chemical purity and isotope enrichment of the $^{29}$Si target are 99.990\% and 99.25\%, respectively.}
    \footnotetext[0]{Chemical purity and isotope enrichment of the $^{30}$Si target are 99.8981\% and 99.64\%, respectively.}
    \footnotetext[0]{Chemical purity of the $^\mathrm{nat}$Si targets are 99.99999\% and 99.+\% for plate and powder form, respectively.}
    \footnotetext[0]{Chemical purity of the $^{27}$Al targets are 99.+\% and 99.999\% for with and without case, respectively.}
    \footnotetext[0]{Chemical purity of the $^\mathrm{nat}$Mg, $^{55}$Mn, and $^\mathrm{nat}$Ag targets are 91.+\%, 99.9+\% and 99.98\%, respectively.}
    \end{ruledtabular}
\end{table*}

%% file: table_err.tex
\begin{table}
    \centering
    \caption{Systematic breakdown of the uncertainty. The units of all values in the table are ns.}
    \label{tab:err}
    \begin{tabular}{cccccccc}
    \hline\hline
    Target & Lifetime &\multicolumn{5}{c}{Propagated uncertainty} & Total\\
           &          & $\tau_\mathrm{nucl}$ & $\tau_\mathrm{Ag}$ & $\tau_\mathrm{CNO}$ & $\tau_\mathrm{BG}$ &Range & \\ \hline
            ${}^{28}$Si &743.9 &2.01 &0.44 &4.61 &0.85 &0.32 &5.0 \\
            ${}^{29}$Si &794.2 &1.85 &0.39 &2.89 &0.91 &0.35 &3.5 \\
            ${}^{30}$Si &859.0 &1.96 &0.43 &1.62 &1.08 &0.16 &2.7 \\
            ${}^\mathrm{nat}$Si (plate) &751.6 &0.77 &- &0.68 &0.20 &0.20 &1.0 \\
            ${}^\mathrm{nat}$Si (in case) &749.8 &2.26 &0.35 &1.40 &0.80 &0.18 &2.8 \\
            ${}^\mathrm{nat}$Mg &1065.5 &1.96 &- &0.39 &0.27 &0.36 &2.0 \\
            ${}^{27}$Al (plate) &863.7 &0.94 &- &0.04 &0.23 &0.10 &1.0 \\
            ${}^{27}$Al (in case) &862.5 &3.19 &0.43 &7.58 &0.98 &0.34 &8.1 \\
            ${}^{55}$Mn &230.3 &3.57 &0.66 &2.54 &0.42 &0.96 &4.5 \\
            \hline\hline
    \end{tabular}
\end{table}

%% file: table_comp.tex
\begin{table}
    \centering
    \caption{Comparisons of capture rate ($\Lambda_\mathrm{cap}$ in unit of $\mu$s$^{-1}$) between experimental value and theoretical calculations.
    The capture rate calculated by Primakoff~\cite{Primakoff1959-rs} and Goulard-Primakoff~\cite{Goulard1974-sr} formulas ($\Lambda_\mathrm{cap}^\mathrm{P}$ and $\Lambda_\mathrm{cap}^\mathrm{GP}$, respectively) together with calculated $Z_\mathrm{eff}$ values are listed.
    For the natural abundance targets, a weighted average of the mass number and $Z_\mathrm{eff}$ by their isotopic aboundance~\cite{natural_abundance} is used to calculate Primakoff's formulas.
    The capture rates calculated by other theories ($\Lambda_\mathrm{cap}^\mathrm{th}$) are also shown.
    }
    \label{tab:comp}
    \begin{ruledtabular}
    \begin{tabular}{cD{.}{.}{7}D{.}{.}{4}D{.}{.}{4}D{.}{.}{4}D{.}{.}{6}}
    \multicolumn{1}{c}{Nucleus}&
    \multicolumn{1}{c}{Exp.}&
    \multicolumn{3}{c}{Primakoff}&
    \multicolumn{1}{c}{Others}
    \\

    &
    \multicolumn{1}{c}{$\Lambda_\mathrm{cap}$}&
    \multicolumn{1}{c}{$Z_\mathrm{eff}$}&
    \multicolumn{1}{c}{$\Lambda_\mathrm{cap}^\mathrm{P}$}&
    \multicolumn{1}{c}{$\Lambda_\mathrm{cap}^\mathrm{GP}$}&
    \multicolumn{1}{c}{$\Lambda_\mathrm{cap}^\mathrm{th}$}
    \\\hline

    $^{28}$Si         & 0.893(9)                   & 12.386 & 0.875 & 0.921 & \multicolumn{1}{c}{0.968~\cite{Chiang1990-vf}}\\
                      &                            &        &       &       & \multicolumn{1}{c}{0.823~\cite{Zinner2006-pe}}\\
                      &                            &        &       &       & \multicolumn{1}{c}{0.789~\cite{Marketin2009-eu}}\\
                      &                            &        &       &       & \multicolumn{1}{c}{0.892~\cite{Giannaka2015-en}}\\
    $^{29}$Si         & 0.808(6)                   & 12.382 & 0.767 & 0.791 & \\
    $^{30}$Si         & 0.713(4)                   & 12.379 & 0.665 & 0.673 & \\
    $^\mathrm{nat}$Si & 0.8794(18)\footnotemark[1] & 12.385\footnotemark[2] & 0.863 & 0.907 & \\ 
    $^\mathrm{nat}$Mg & 0.4856(18)                 & 10.837\footnotemark[2] & 0.489 & 0.511 & \multicolumn{1}{c}{0.612~\cite{Chiang1990-vf}}\\
                      &                            &        &       &       & \multicolumn{1}{c}{0.454~\cite{Zinner2006-pe}}\\
    $^{27}$Al         & 0.7059(13)\footnotemark[1] & 11.622 & 0.589 & 0.606 & \multicolumn{1}{c}{0.698~\cite{Chiang1990-vf}}\\
    $^{55}$Mn         & 3.90(8)                    & 19.379 & 3.542 & 3.567 & \\ 
    $^\mathrm{nat}$Ag & 11.07(7)                   & 27.883\footnotemark[2] & 12.095 & 12.117 & \\ 
    \end{tabular}
    \footnotetext[1]{Average of two experimental data in Table~\ref{tab:result}.}
    \footnotetext[2]{Weighted average of $Z_\mathrm{eff}$ by the natual abundance of each isotopes~\cite{natural_abundance}.}
    \end{ruledtabular}
\end{table}